\documentclass[12pt,preprint]{aastex}

\newcommand{\etal}{{et al.}}
\newcommand{\eg}{{e.g.,}}

\begin{document}

\title{An Infrared Comparison of Type-1 and Type-2 Quasars}

\author{Kyle D. Hiner\altaffilmark{1}, Gabriela
  Canalizo\altaffilmark{1}, Mark Lacy\altaffilmark{2},  Anna
  Sajina\altaffilmark{3}, Lee Armus\altaffilmark{2}, Susan
  Ridgway\altaffilmark{4}, Lisa Storrie-Lombardi\altaffilmark{2}} 

\altaffiltext{1}{Department of Physics and Astronomy \& Institute of
  Geophysics and Planetary Physics, University of California,
  Riverside, CA 92521, USA; email: kyle.hiner@email.ucr.edu,
  gabriela.canalizo@ucr.edu}  

\altaffiltext{2}{Spitzer Science Center, California Institute of
  Technology, Pasadena, CA 91125, USA; email: mlacy@ipac.caltech.edu,
  lee@ipac.caltech.edu, lisa@ipac.caltech.edu} 

\altaffiltext{3}{Department of Physics \& Astronomy, Haverford College, Haverford, PA 19041, USA; email: asajina@haverford.edu}

\altaffiltext{4}{NOAO/CTIO, 950 N. Cherry Ave, Tuscon, AZ 85719, USA;
  email: seridgway@ctio.noao.edu}

\begin{abstract}
We model the optical to far-infrared SEDs of a sample of six type-1 and six type-2 quasars selected in the mid-infrared. The objects in our sample are matched in mid-IR luminosity and selected based on their {\it Spitzer} IRAC colors. We obtained new targeted {\it Spitzer} IRS and MIPS observations and used archival photometry to examine the optical to far-IR SEDs. We investigate whether the observed differences between samples are consistent with orientation-based unification schemes. The type-1 objects show significant emission at 3 \micron. They do not show strong PAH emission and have less far-IR emission on average when compared to the type-2 objects. The SEDs of the type-2 objects show a wide assortment of silicate features, ranging from weak emission to deep silicate absorption. Some also show strong PAH features. In comparison, silicate is only seen in emission in the type-1 objects. This is consistent with some of the type-2s being reddened by a foreground screen of cooler dust, perhaps in the host galaxy itself. We investigate the AGN contribution to the far-IR emission and find it to be significant. We also estimate the star formation rate for each of the objects by integrating the modeled far-IR flux and compare this with the SFR found from PAH emission. We find the type-2 quasars have a higher average SFR than the type-1 quasars based on both methods, though this could be due to differences in bolometric luminosities of the objects. While we find pronounced differences between the two types of objects, none of them are inconsistent with orientation-based unification shemes.
\end{abstract}

\singlespace
\section{Introduction}
The study of various types
of AGN can provide insight into their origin and evolution. AGN are
classified according to their optical emission lines as type-1 or
type-2. Type-1 AGN have an \textquotedblleft
unobscured\textquotedblright~view of the hot accretion disk and broad
emission lines originating from fast moving clouds that surround a
supermassive black hole. Type-2 objects are known as \textquotedblleft
obscured\textquotedblright~AGN, because no broad lines or continuum
emission from the accretion disk are observed, yet the spectra still
show narrow emission lines similar to those of the type-1 objects. 

The interpretation of the differences between type-1 and type-2
quasars (the most luminous AGN) remains an open issue. The
orientation-based unification scheme proposes the two types of objects
are intrinsically the same, but appear different observationally due
to the presence of an obscuring torus of dust close to the central
power source (see Antonucci 1993). Type-1 objects are seen from an
orientation which allows an unobscured view of the central
source. This allows the broad line region of the quasar to be
observed. However, type-2 objects are seen through the dusty torus
material, which hides the central continuum source and broad-line
region. Evidence that this model is at least partly correct comes from
observations of polarized broad-line emission in galaxies classified
as type-2 quasars (\eg~Zakamska et al. 2005). But this model does not
provide insight into the origin of quasar activity. 

An alternative model proposes that quasar activity is triggered
by galactic mergers and evolves through an obscured stage
\citep{S88}. As the galaxies merge, gas and dust are stirred up,
triggering quasar activity and star formation. Initially the quasar
would be obscured by the dust, but over time will become unobscured as
dust and gas are blown away from the nuclear region by quasar and/or
supernovae driven winds and radiation pressure. Indeed, several host
galaxy studies have been made which show that luminous AGN are hosted
by massive, mainly spheroid-dominated galaxies, with a significant
fraction, but not all, showing signs of a relatively recent merger and
associated star formation (\eg~Dunlop et al. 2003; Canalizo et
al. 2007; Bennert et al. 2008). Thus the evolution model predicts that
the host galaxies of type-2 quasars should have more recent star
forming activity than those of type-1s. Although the discovery of
polarized broad lines is a powerful argument in favor of the
orientation-based model, it does not rule out an evolutionary link.

The obscuring dust in a host galaxy would be apparent in its
observed SED. Dust close to the AGN would be heated by the central power
source and would thermally re-radiate in the mid-IR (100 - 1000K),
dominating the rest-frame SED from $\sim$2 - 40 \micron. This high
temperature dust emission makes the mid-IR SEDs of both
optically-obscured and optically-unobscured AGN quite distinct from
those of either normal or starburst galaxies, whose dust temperatures
are always $\leq$ 100K. By modeling the various components of the SEDs,
we can compare the relative amounts of ``cool'', \textquotedblleft
warm\textquotedblright, and \textquotedblleft
hot\textquotedblright~dust emission present in the host galaxies. This
will give us some idea of what mechanism dominates the heating of the
dust (star formation or AGN), and how much of a role orientation might
play in the observed SEDs. 

In this paper, we present modeled fits to the SEDs of 
a sample of type-1 and type-2 quasars of similar redshift and
luminosity. \citet{Lacy07} presented preliminary fits to the type-2
quasars in this paper. We also measure the objects' far-IR luminosities, PAH luminosities, and estimate their star formation rates. We then compare the results with starburst galaxies. We describe the sample selection in \S2, and the
observations and reductions in \S3. In \S4 we present the analysis and
results including our modeling procedure and resultant spectra, a color-color diagram, and star formation rate (SFR) calculations. Finally, we summarize our results in \S5. Throughout this paper
we adopt the cosmological parameters of H$_{0}$=71 km s$^{-1}$
Mpc$^{-1}$, $\Omega_{M}$=0.27 and $\Omega_{\Lambda}$=0.73.

\section{Sample Selection}
Quasar surveys are now highly effective at finding quasars on the
basis of their optical colors, provided they are not reddened or only
lightly reddened by dust. With surveys such as the Sloan Digital Sky
Survey (SDSS) (Schneider et al. 2003), 2dF (Croom et al. 2004), and
Combo-17 (Wolf et al. 2003), a census of such objects across the full
range of quasar luminosities and redshifts out to z $\sim$ 6 is
largely complete. Only recently have radio-quiet type-2 quasars been
found in substantial numbers. However, there is still no consensus on
how the dust-obscured quasars are related to the unobscured
population. In particular, it is difficult to select matched samples
of type-1 and type-2 objects with which to perform such studies. By
using {\it Spitzer} data from the extragalactic First Look Survey (XFLS), we
have been able to construct samples of type-1 and type-2 quasars
matched in mid-IR luminosity and redshift.

\citet{Lacy04} showed that quasars occupy a distinct \textquotedblleft
sequence\textquotedblright~of a {\it Spitzer} IRAC color-color plot
(their Fig.~1). Follow-up spectroscopy of a sample of mid-IR selected AGN
and quasar candidates showed that about 2/3 of AGN selected in this
manner may be sufficiently obscured by dust in the optical to be missing from 
optically-selected quasar samples \citep{Lacy07}. 
We selected six type-2 quasars and six type-1 quasars (see
Table~\ref{targets}) from the 1 mJy  8 $\mu$m flux-limited sample of
\citet{Lacy04}, with similar mid-IR colors and 5.8 \micron~luminosities. These
objects are approximately matched in  mid-IR luminosity. Our modeling
(see \S\ref{Modeling}) shows they have mid-IR power-law luminosities
within a factor of five from each other. Furthermore, the observed frame 24
\micron~luminosities of the type-1 and type-2 quasars are not
statistically different from each other. It is important that the
sample be matched in luminosity, since the quasar luminosity will
affect the rate at which dust and gas is removed from the nuclear
region. 

\section{Observations and Data Reduction}
We supplement data from the {\it Spitzer} Extragalactic First Look
Survey (xFLS) \citep{Fadda06,Frayer06} with targeted
observations in the 70 \micron~and 160 \micron~using the Multiband Imaging
Photometer for {\it Spitzer} (MIPS). Each of our targets was detected at
the 24 \micron~wavelength in the {\it Spitzer} xFLS. Five of our
targets were detected in the 70 \micron~xFLS mosaic, and we
measured the 70 \micron~fluxes for two other targets directly from the
xFLS mosaic image. The remaining five targets of our sample were not
detected in the 70 \micron~xFLS mosaic and so required targeted
observations. We observed each of these targets for 500 seconds. All
of the targets were observed with the MIPS 160 \micron~band for 2000
seconds each. 

The photometric data were reduced through the standard {\it Spitzer} MIPS
pipeline. We used the post basic calibrated (post-BCD) filtered mosaic
images to measure the flux of each target. The pixel scales of the
post-BCD MIPS images are 4 arcsec pixel\textsuperscript{-1} at
70 \micron~and 8 arcsec pixel\textsuperscript{-1} at 160 \micron~.

We also observed each target using the short-low and long-low
resolution modes with the {\it Spitzer} infrared spectrograph
(IRS). The spectra were reduced through the standard {\it Spitzer}
pipeline. We manually masked the bad pixels in the background
subtracted spectra (difference between nodded exposure positions)
using the IDL program
IRSCLEAN\footnote{http://ssc.spitzer.caltech.edu/archanaly/contributed/irsclean}. 
We then extracted the spectra using the standard point source extraction in the {\it Spitzer}
SPICE\footnote{http://ssc.spitzer.caltech.edu/postbcd/spice.html} software
tool, and combined the individual spectra to create a continuous mid-IR spectrum
for each target. Lastly we included photometry measured from our
MIPS data (Table \ref{targets}) and data available for our targets from SDSS, 2MASS and {\it
  Spitzer} IRAC observations (Table \ref{mphot}) to produce an observed frame SED that
ranges from optical through the infrared. 

\subsection{Photometry}
We used the DAOPHOT package in IRAF to measure the 70 \micron~and
160 \micron~fluxes of our targets. For the 70 \micron~measurements we
examined each image individually and determined the best aperture and
sky annulus that maximized the signal-to-noise ratio. We then
applied aperture corrections from the {\it Spitzer} Space Telescope MIPS Data
Handbook\footnote{http://ssc.sptizer.caltech.edu/mips/apercorr/}, and
compared the measured flux to a 3$\sigma$ limit obtained by measuring
the noise in the image.

Each of our 12 targets was observed in the 160 \micron~MIPS band,
though many of our targets were not detected at this wavelength. We chose an
example image with a strong detection and determined the aperture and sky annulus sizes that maximized the signal-to-noise ratio. We used those
parameters while measuring the other images with bright detections. 
The same procedure was used for the images with targets that were not
detected. We then compared the measured fluxes of each
of the targets to the 3$\sigma$ limit determined from the noise in the
image. Five of the 12 targets have fluxes less than the 2$\sigma$
noise limit, and we count them as non-detections. Five other targets
have fluxes $>$2$\sigma$, and only one has 160 \micron~flux
$>$3$\sigma$. The remaining target is discussed following.

The target SDSS 1711+5855 required special attention,
because there was a bright object near the target. We fit a PSF using
several other bright objects in the image, then subtracted sources
from the image leaving only our target. We then measured the flux of
SDSS 1711+5855 from the subtracted image. This procedure was successful at 70 \micron, but our attempts at
removing the extra source at 160 \micron~resulted in an
image with significant residuals. Therefore, we only report an upper limit measured from the combined flux of the objects at 160 \micron. This upper limit was not used in our modeling procedure and analysis.

Table~\ref{targets} shows the results of our MIPS flux measurements
combined with data from the xFLS.

\begin{deluxetable}{ccccccccc}
\tablecolumns{9}
\tablecaption{Targets and their Fluxes in mJy}
\tablehead{
\colhead{Target} & \colhead{Type} & \colhead{z} & \colhead{$\emph{f}_{24}$} & \colhead{$\sigma_{24}$} & \colhead{$\emph{f}_{70}$} & \colhead{$\sigma_{70}$} & \colhead{$\emph{f}_{160}$} & \colhead{$\sigma_{160}$}}
\startdata
SDSS 171117.66+584123.8 & 1 & 0.617 & 5.77 & 0.07 & 34.9 & 6.3 & 30.4 & 13.4 \\
SDSS 171126.94+585544.2 & 1 & 0.537 & 3.45 & 0.07 & 12.7 & 6.6 & $<$102.1 & \nodata \\
SDSS 171334.03+595028.3 & 1 & 0.615 & 5.38 & 0.07 & 7.5 & 1.6 & $<$11.1 & 3.7 \\
SDSS 171736.91+593011.5 & 1 & 0.599 & 6.38 & 0.05 & 12.9 & 2.5 & $<$15.9 & 5.3 \\
SDSS 171748.43+594820.6 & 1 & 0.763 & 3.04 & 0.04 & 9.7 & 4.1 & $<$9.6 & 3.2 \\
SDSS 171818.14+584905.2 & 1 & 0.634 & 4.06 & 0.06 & 35.0 & 6.1 & 28.9 & 13.9 \\
SST 171106.8+590436 & 2 & 0.462 & 3.65 & 0.06 & 26.3 & 4.9 & $<$19.5 & 6.5 \\
SST 171147.4+585839 & 2 & 0.800 & 4.84 & 0.07 & 18.8 & 5.7 & 31.9 & 13.7 \\
SST 171324.1+585549 & 2 & 0.609 & 4.94 & 0.07 & 16.1 & 1.5 & 33.8 & 10.7 \\
SST 171831.7+595317 & 2 & 0.700 & 8.27 & 0.04 & 27.1 & 4.4 & 31.2 & 12.8 \\
SST 172123.1+601214 & 2 & 0.325 & 13.34 & 0.07 & 12.0 & 6.2 & $<$9.3 & 3.1 \\
SST 172458.3+591545 & 2 & 0.494 & 2.60 & 0.06 & 24.6 & 4.9 & 38.1 & 15.0 \\
\enddata
\label{targets}
\end{deluxetable}

\subsection{Spectra}
\begin{figure}
\epsscale{1}
\plottwo{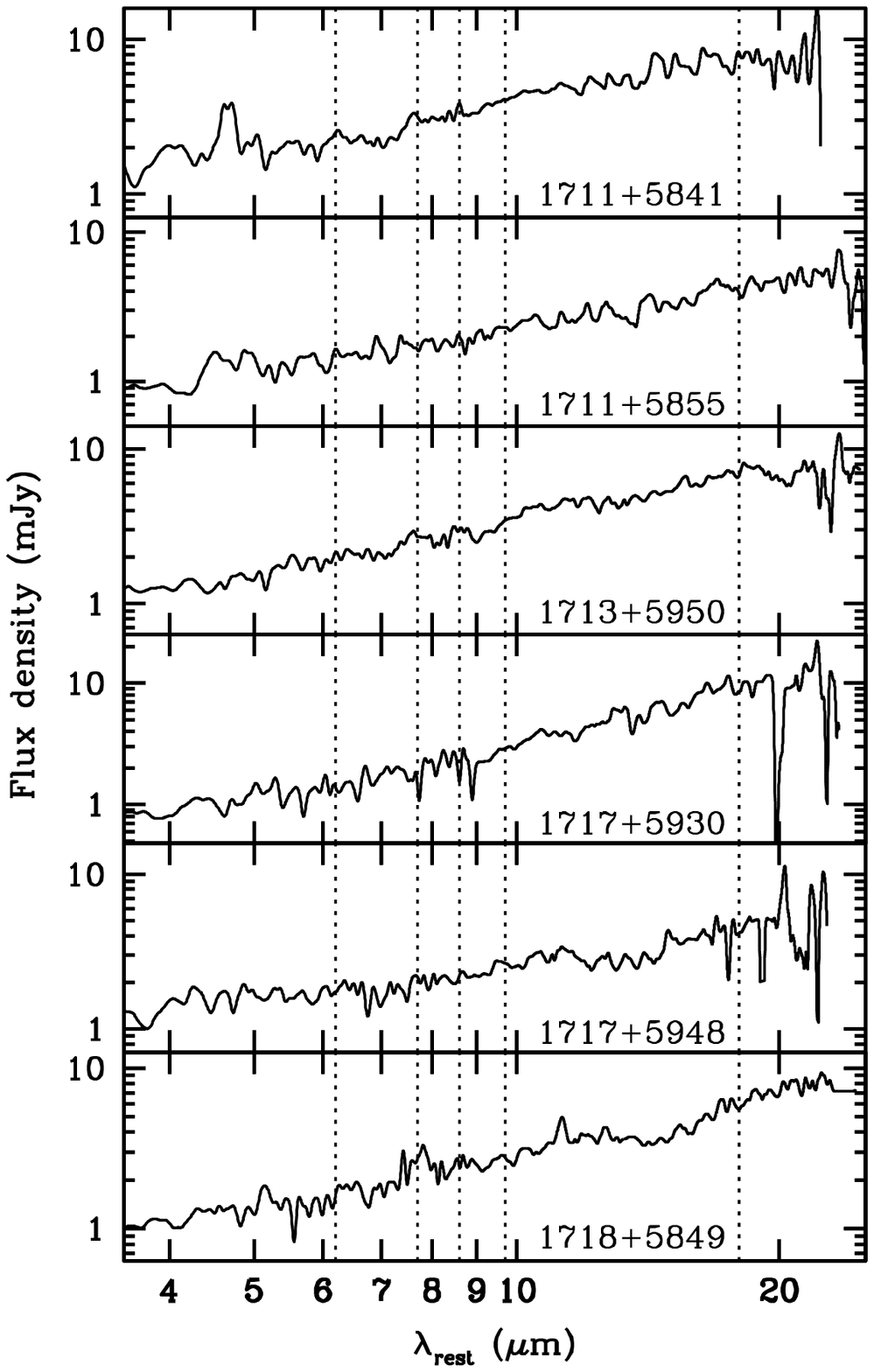}{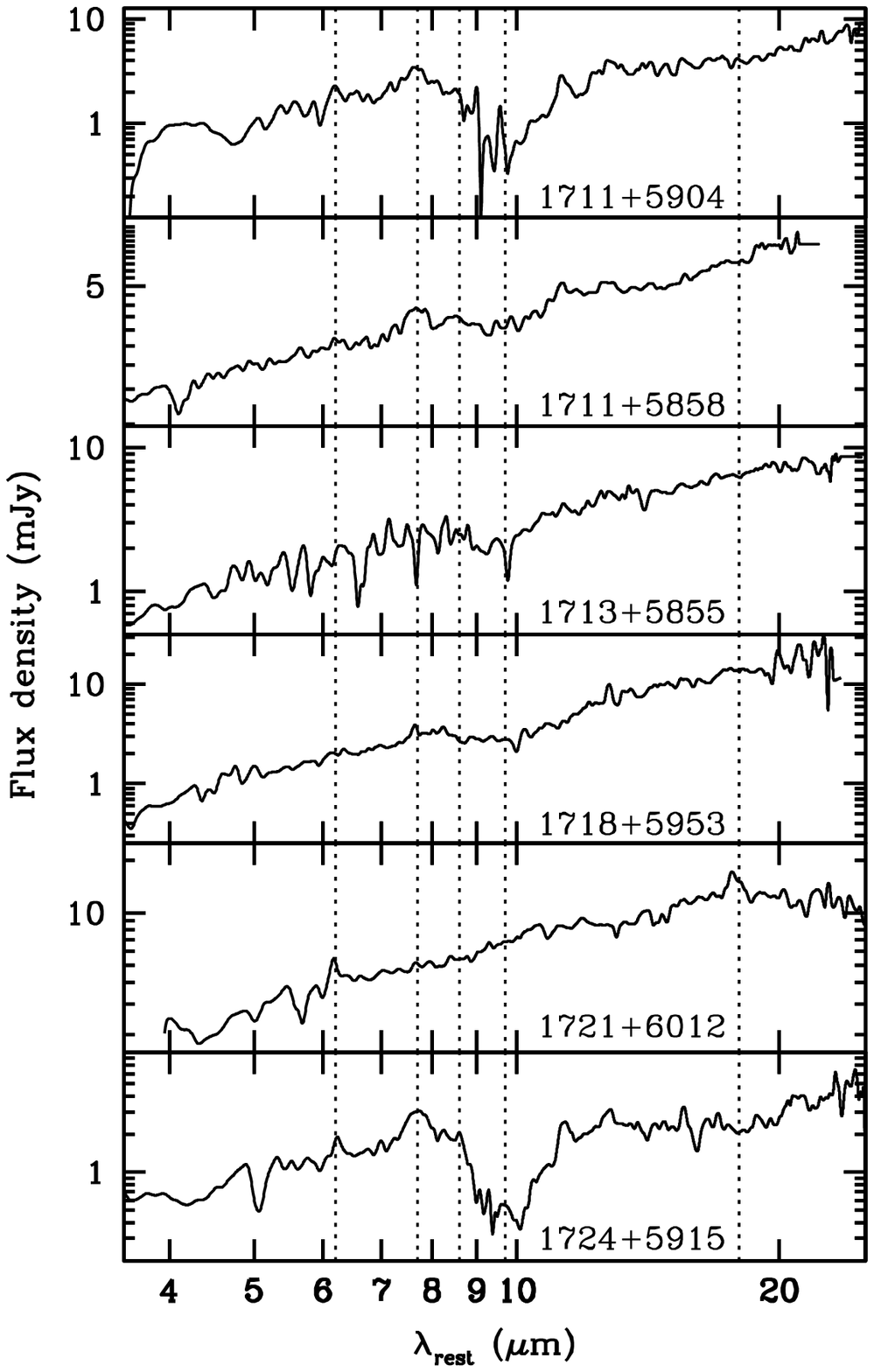}
\caption{{\it Left}: IRS spectra for our type-1 quasar sample. {\it
    Right}: IRS spectra for our type-2 quasar sample. The spectra have
been smoothed with a gaussian with sigma of five pixels. Vertical
dotted lines indicate the wavelengths for 6.2, 7.7 and 8.6 \micron~PAH
features, and 9.7 and 18 \micron~silicate features.}
\label{panel}
\end{figure}

We present the {\it Spitzer} IRS spectra for our sample in
Fig.~\ref{panel}. The spectra show distinct differences. The spectra of the type-1 quasars tend to have fewer features and are more homogeneous than those of the type-2s.  The type-1 objects show significant continuum emission near 3 \micron, little PAH emission, and no significant silicate absorption. The type-2
spectra are more heterogenous, showing a variety of silicate
absorption depths and PAH strengths, consistent with previous observations of AGN (\eg~Hao \etal~2007; Spoon \etal~2007; Zakamska \etal~2008) as well as starburst galaxies, ULIRGs and sub-mm galaxies (\eg~Brandl \etal~2006; Sajina \etal~2007; Lutz \etal~2008; Menendez-Delmestre \etal~2009). The type-2 quasars have on average double the PAH luminosity of the type-1 quasars (see Table \ref{lum}).

\citet{Spoon07} classify spectra
based on their 6.2 \micron~PAH equivalent width and 9.7
\micron~silicate absorption strength. Our type-2 spectra are similar
to their 1B, 1C and 2B classifications. These classifications are
dominated by Seyferts, starbursts, and ULIRGs, respectively. Our
type-1 quasars have spectra most similar to the 1A classification of
galaxies by Spoon \etal

The variation of silicate absorption strengths in the
type-2 spectra of our sample is also consistent with that seen by
\citet{Z08}, who study a sample of optically selected type-2 quasars
from the SDSS. Zakamska \etal~find silicate absorption similar to some
of our sample, including two objects that have more significant
absorption than our sample (see Lacy \etal~2007). Four of their
objects appear to have no silicate absorption, including one that may
show silicate emission, consistent with our mid-IR selected type-1
quasars. They also address the correlation of silicate absorption with
axial ratio found by \citet{Lacy07}. Lacy \etal~interpreted this correlation to indicate that the silicate dust resides in the host galaxy, and not the nuclear region. The objects of \citet{Z08} show no correlation,
and they propose this could be a difference in samples due to
selection biases based on luminosity.

\section{Analysis and Results}
\subsection{Modeling}
\label{Modeling}
We modeled the SEDs of our sample from the optical out to the
far-IR. We used available SDSS and 2MASS photometry, IRAC bands 1 and 2
(3.6 \micron~and 4.5 \micron, respectively) photometry \citep{Lacy05}
and our {\it Spitzer} IRS mid-IR spectra and MIPS photometry. This additional photometry used in the modeling procedure is presented in Table \ref{mphot}. The
modeling code is described by \citet{S06}. Modifications
to the code to enable modeling of the type-2
objects in our sample were briefly described by \citet{Lacy07}. 
Here we provide a fuller description of the modeling, and the additional
modifications required to fit the type-1 objects. The modeling is
phenomenological in nature due to the current lack of a good physical
understanding of the nature and geometry of the dust emitting regions,
particularly in the mid-IR, thus many of the parameters do not have a
direct physical interpretation. Nevertheless, we considered the effort
worthwhile as it allows an empirical breakdown of the contributions of
starlight, hot dust emission from the AGN, and cooler dust emission.
 
The type-2 SEDs were modeled using the following components. (1) A single
stellar population from \citet{BC03} with an age of 5 Gyr to 
represent the evolved stellar population. (2) A power-law component for the
mid-IR emission, with an exponential cutoff at short wavelengths
to represent dust sublimation, and a Fermi function cutoff at long wavelengths
designed to match the SED shape of the least star-forming object in the 
sample of type-2s, SST 1721+6012. The functional form used was:

\begin{equation}
 L_{\rm AGN}=\frac{L^{0}_{\rm AGN}\nu^{1-\alpha} e^{h \nu / k_B T_{\rm sub}}}{e^{(\nu - \nu_{\rm hcut})/w}+1}
\end{equation}

where the normalization, $L^0_{\rm AGN}$, 
power-law index $\alpha$ and $T_{\rm sub}$ (a proxy for the sublimation temperature) are 
allowed to vary, and $\nu_{\rm hcut}$ and $w$ were fixed at $0.11\times 10^{14}$Hz and 
$0.017\times 10^{14}$Hz, respectively ($h$ and $k_{\rm B}$ are the Planck and Boltzman
constants, respectively). The Galactic Center 
extinction curve of \citet{CT06} was then applied to this 
continuum to fit any silicate absorption feature. (3) A warm dust component to represent
the small grain emission from H{\sc ii} regions, represented by
a power-law with fixed high and low frequency cutoffs:

\begin{equation}
L_{\rm SG} = L^{0}_{\rm SG}\nu^{1-\gamma} e^{-\nu/\nu_{\rm sgl}} e^{-\nu_{\rm sgh}/\nu}
\label{SGeq}
\end{equation}

This component is poorly constrained in most of the fits, so $\gamma$ was fixed at a 
typical value of two (e.g.\ Sajina et al.\ 2006). The high and low frequency 
cutoffs, $\nu_{\rm sgl}$ and $\nu_{\rm sgh}$ were set to $0.062\times 10^{14}$Hz and
$0.17\times 10^{14}$Hz, respectively. (4) a thermal modified blackbody model for the far-infrared emission:

\begin{equation}
L_{\rm FIR} = \frac{L^0_{\rm FIR} \nu^{3+\beta}}{e^{h\nu/k_{\rm B}T}-1}
\label{FIReq}
\end{equation}

For most objects, we fixed the temperature of this component, $T$, to 45K,
because of the uncertainty in our 160 \micron~photometry. $\beta$ was fixed at
1.5. In addition, the PAH model described in \citet{Lacy07} was
added when evidence of PAH emission was seen, and SST 1721+6012 had a silicate 
emission profile added, whose shape was approximated by that of the
extinction curve.

For the type-1 quasars, component (1) was replaced by an SDSS
composite quasar spectrum. As the composite of \citet{VB01}~suffers from noticeable host galaxy contamination at the long wavelength
end, we constructed a new composite by subtracting the continuum
from the composite of Vanden Berk et al., and adding the residual emission
line composite to a new continuum constructed using line-free SED points
from the composite of \citet{R06}. 

Most type-1 objects also required 
an extra near-infrared component to match
the SEDs. This ``very hot'' component was modeled as a 1000 K modified blackbody
(Eq.~\ref{FIReq}~with $T=1000$). The physical origin of this component
is unclear, but it is most likely due to hot dust close to the 
sublimation radius (e.g.\ Netzer et al.\ 2007). \citet{G07} show that
the strength of this component correlates with quasar luminosity. We
do not find this to be true for our sample, but note that \citet{G07}
have a much larger sample size spread over a range of
luminosities.

\begin{figure}
\epsscale{1}
\plotone{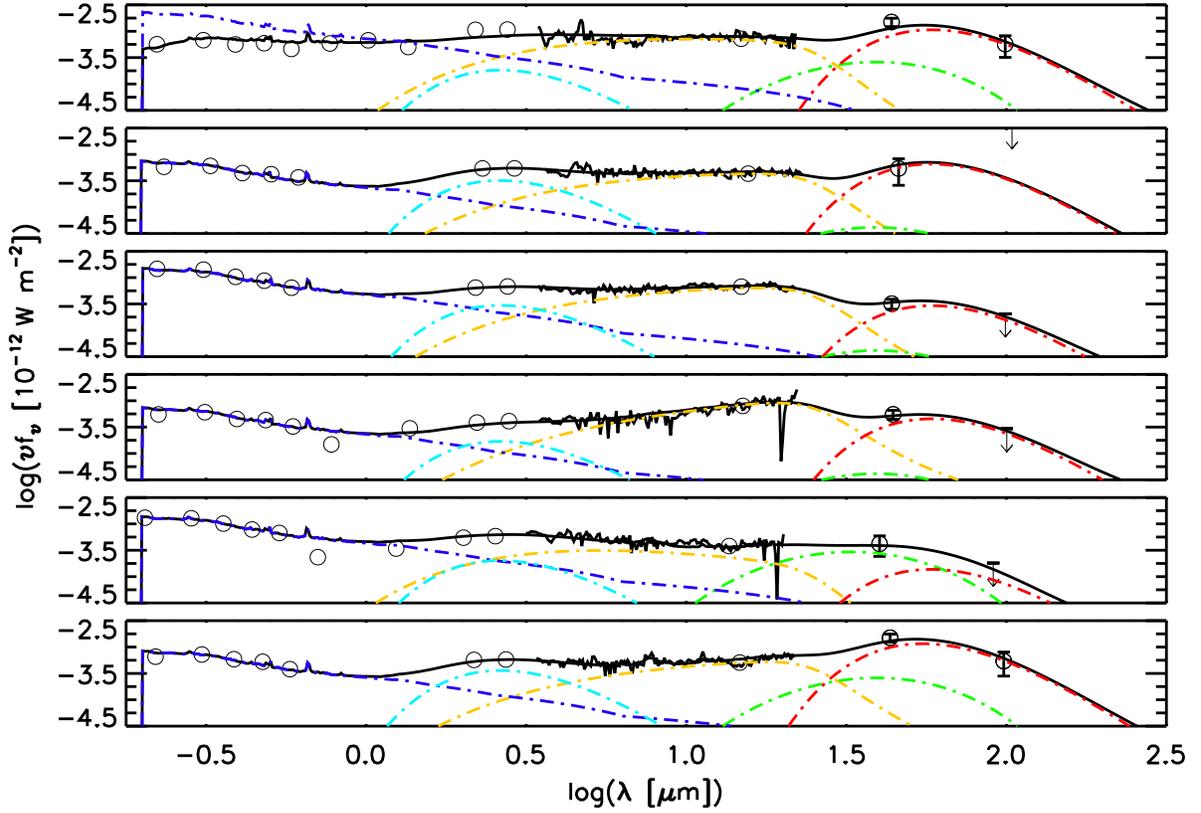}
\caption{The type-1 sample. The spectra are presented in their
  rest frame. From top to bottom are SDSS 1711+5841, 1711+5855, 1713+5950,
  1717+5930, 1717+5948, and 1718+5849. The photometry measurements are
  plotted as circles and the IRS data are plotted as a solid line. We
  show error bars for the MIPS 70 \micron~and 160
  \micron~photometry. Error bars on other photometry measurements are
  smaller than the plotting symbol. The individual components of the
  model are plotted as dot-dashed lines and the overall fit as the
  solid black line. The blue line represents the quasar composite component, the cyan line the 1000 K component, the orange line the mid-IR power-law, the green line the warm small grain dust power-law, and the red line the cool 45 K modified blackbody.} 
\label{type1s}
\end{figure}

\begin{figure}
\epsscale{1}
\plotone{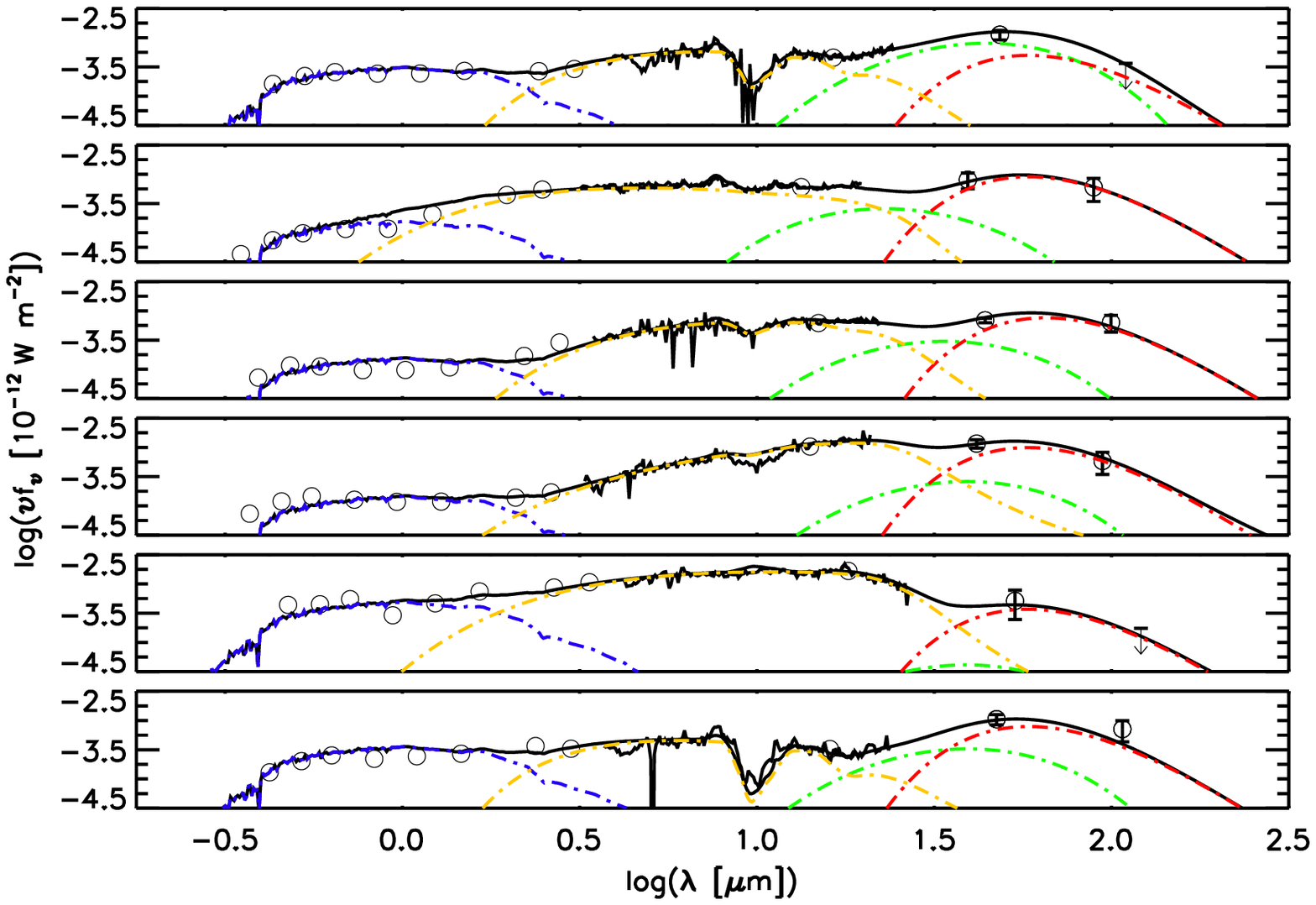}
\caption{The type-2 sample. The spectra are presented in their
  rest frame. From top to bottom are SST  1711+5904, 1711+5858,
  1713+5855, 1718+5953, 1721+6012, 1724+5915. The plotted symbols are
  the same as in Fig.~\ref{type1s}. The photometry measurements are
  plotted as circles and the IRS data are plotted as a solid line. We
  show error bars for the MIPS 70 \micron~and 160
  \micron~photometry. Error bars on other photometry measurements are
  smaller than the plotting symbol. The individual components of the
  model are plotted as dot-dashed lines and the overall fit as the
  solid black line. The blue line represents the host galaxy stellar component, the orange line the mid-IR power-law, the green line the warm small grain dust power-law, and the red line the cool 45 K modified blackbody.}
\label{type2s}
\end{figure}

The modeling results for the type-1 objects are presented in
Fig.~\ref{type1s}, and the results for
the type-2 objects are presented in Fig.~\ref{type2s}. Several
differences between the type-1 sample and the type-2 sample are
immediately apparent from the fitting results. The type-1 objects have a strong optical quasar component that can extend significantly into the mid-IR. The 1000 K component is significant in all six type-1s and also contributes to the mid-IR. Three of the six type-1 object SEDs showed no evidence of the warm small grain dust power-law emission. The type-2 object SEDs are dominated in the mid-IR by hot dust with no contribution from the optical stellar light and no evidence for the 1000 K component. Also present in the mid-IR portion of the type-2 objects is a variety of PAH emission and silicate absorption. Five of the six type-2 objects have significant contributions to the mid and far-IR from the warm small grain dust power-law. SST 1721+6012 was the only type-2 not fit well with this component.

We note that the results of the fitting procedure for the object SST 1724+5915 underestimate both the observed amount of PAH emission and the far-IR emission. This is likely due to the large amount of silicate absorption also present in the spectrum of the object. In particular the 9.7 \micron~absorption feature can affect the shape of the 7.7 \micron~PAH feature, making it difficult to fit with an empirical template. We include the data in our subsequent analysis, but note that star formation rates derived from the far-IR emission and the PAH emission likely underestimate the true star formation of this object (see \S\ref{SFRs}).

\begin{figure}
\epsscale{1}
\plotone{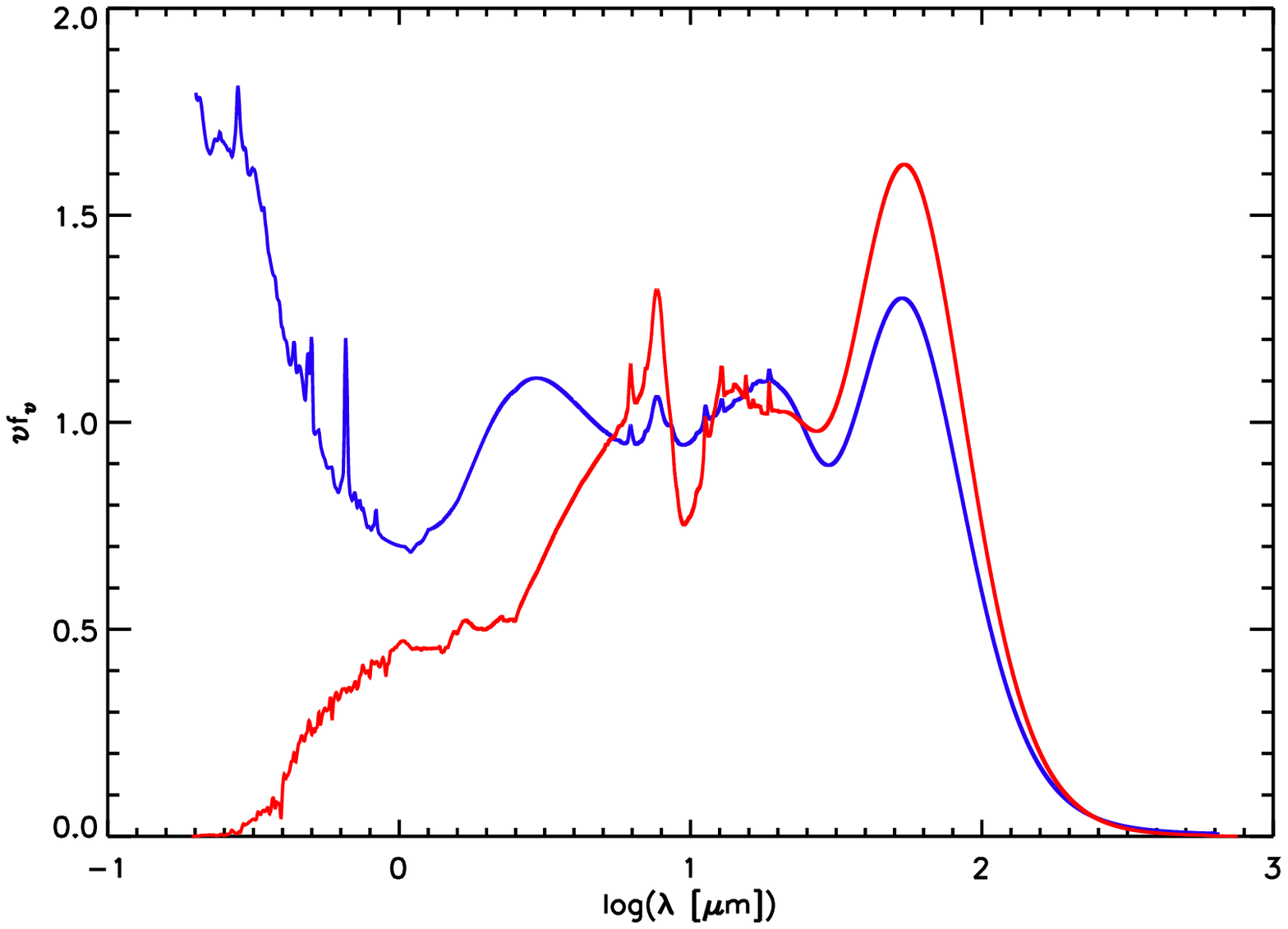}
\caption{Averaged modeled SEDs of type-1 (blue) and type-2 (red) quasars. The spectra are normalized at 24 \micron. We plot $\nu$f$_{\nu}$ to highlight the differences between the two.}
\label{average}
\end{figure}

We averaged the modeled type-1 SEDs and type-2 SEDs to highlight the differences between the two samples. We first normalized the SED of each target at 24 \micron~to ensure each object had equal weight. By chosing to normalize at 24 \micron~we can compare the relative amounts of hot and cool dust in the averaged SEDs. Though at 24 \micron, extinction is still present, it is significantly reduced compared to shorter wavelengths. It is clear from Fig.~\ref{average} that on average type-2 objects show more PAH emission, more silicate absorption and more far-IR emission than type-1 objects. In addition, while the very hot 3 \micron~bump is clear in the type-1 spectrum, no bump is seen in the type-2 spectrum. We note the caveat that the type-1 and type-2 objects could have equivalent amounts of PAH emission if that emission were significantly diluted by a strong mid-IR continuum in type-1 objects. 

The important question is if these observed differences could arise solely due to orientation effects. In that scenario the optical through mid-IR light is absorbed and re-radiated in the far-IR. To test this, we integrated the optical through far-IR fluxes of the normalized and averaged spectra and found them to be identical. These results are consistent with the orientation hypothesis, though the geometry and distribution of the obscuring dust cannot be proved from this simple test.

Furthermore, under the orientation hypothesis, the dust tori of the type-1 and type-2 quasars should have the same intrinsic luminosities. Our modeling results show that the tori of the type-2 quasars are slightly more luminous than the tori of the type-1 quasars, though the difference is not statistically significant. This may be a result of our mid-IR selection technique, as those wavelengths are still somewhat affected by extinction. In the next section we examine the PAH emission and the star formation contribution to the far-IR emission to gain more insight into whether the far-IR emission is different between the two objects.

\begin{deluxetable}{ccccccccccc}
\tablewidth{6in}
\tabletypesize{\scriptsize}
\tablecaption{Additional Photometry Used in Modeling Procedure}
\tablehead{\colhead{Target} & \colhead{{\it u}} & \colhead{{\it g}} & \colhead{{\it r}} & \colhead{{\it i}} & \colhead{{\it z}} & \colhead{J} & \colhead{H} & \colhead{K} & \colhead{S$_{3.6}$} & \colhead{S$_{4.5}$}}
\startdata
SDSS 171117.66+584123.8 & 19.23 & 18.70 & 18.61 & 18.33 & 18.37 & 17.04 & 16.10 & 15.63 & 1.26 & 1.61\\
SDSS 171126.94+585544.2 & 19.27 & 18.86 & 18.94 & 18.75 & 18.70 & \nodata & \nodata & \nodata & 0.64 & 0.81\\
SDSS 171334.03+595028.3 & 18.24 & 17.95 & 18.04 & 17.98 & 18.14 & \nodata & \nodata & \nodata & 0.76 & 1.02\\
SDSS 171736.91+593011.5 & 19.36 & 18.85 & 18.92 & 18.72 & 18.82 & 18.53 & \nodata & 16.19 & 0.46 & 0.61 \\
SDSS 171748.43+594820.6 & 18.42 & 18.06 & 18.07 & 18.14 & 18.09 & 18.04 & \nodata & 16.05 & 0.65 & 0.90\\
SDSS 171818.14+584905.2 & 19.17 & 18.70 & 18.70 & 18.57 & 18.70 & \nodata & \nodata & \nodata & 0.68 & 0.87\\
SST 171106.8+590436 & 22.30 & 21.52 & 20.07 & 19.51 & 19.15 & 18.0 & 17.2 & 16.3 & 0.32 & 0.44\\
SST 171147.4+585839 & 22.94 & 22.30 & 21.61 & 20.80 & 20.31 & 18.8 & 18.0 & 16.6 & 0.52 & 0.82\\
SST 171324.1+585549 & 22.82 & 21.83 & 20.95 & 20.21 & 20.06 & 19.0 & 18.2 & 17.3 & 0.20 & 0.43\\
SST 171831.7+595317 & 22.93 & 21.91 & 20.93 & 20.18 & 19.78 & 18.7 & 18.0 & 17.2 & 0.16 & 0.26\\
SST 172123.1+601214 & 21.69 & 20.28 & 18.99 & 18.74 & 18.33 & 17.8 & 16.5 & 15.2 & 1.03 & 1.60\\
SST 172458.3+591545 & 21.94 & 21.69 & 20.31 & 19.61 & 19.17 & 18.1 & 17.2 & 16.3 & 0.43 & 0.49\\
\enddata
\tablecomments{Additional photometry for the modeling procedure was retreived from NASA Extragalactic Database (NED). SDSS {\it u}, {\it g}, {\it r}, {\it i}, and {\it z} magnitudes are listed in the AB system \citep{O90}. The {\it u} and {\it g} photometry were not used in the fitting of the type-2 targets. 2MASS J, H and K$_{s}$ magnitudes are presented in Vega system. IRAC bands 1 and 2 (3.6 \micron~and 4.5 \micron) are listed in mJy.}
\label{mphot}
\end{deluxetable}

\begin{figure}
\epsscale{1}
\plotone{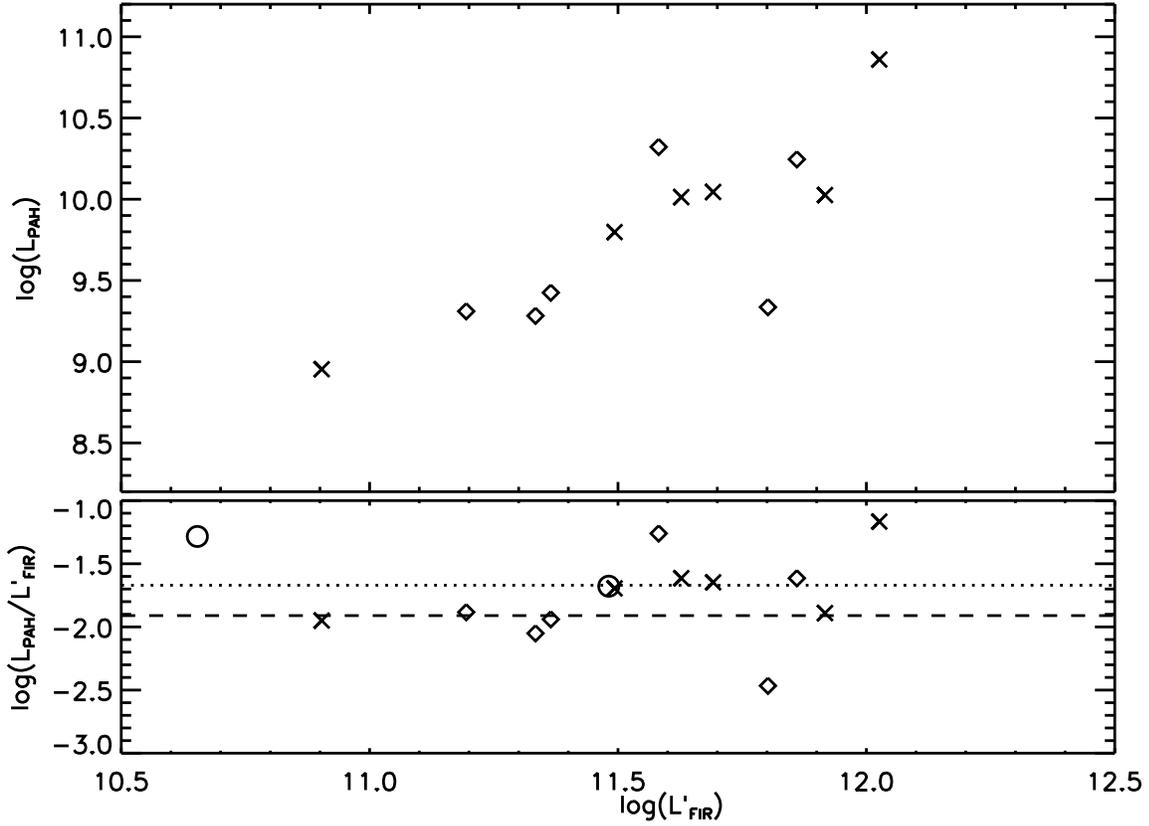}
\caption{{\it Top}: Integrated PAH luminosity as measured from our scaled template is plotted against far-IR luminosity. {\it Bottom}: Relative amount of PAH emission compared to the far-IR dust emission. In both plots, L$'_{FIR}$ was measured from the warm and cold far-IR components of the model only. L$_{PAH}$ and L$'_{FIR}$ are in L$_{\odot}$ units. Type-1 quasars are plotted as diamonds, and type-2 quasars are plotted as Xs. The starburst galaxies NGC 7714 and NGC 2623 are plotted as circles \citep{M07}. The dashed line indicates the median ratio of the type-1 objects of our sample, and the dotted line indicates the median ratio of the type-2 objects.}
\label{PAHFIR}
\end{figure}

\subsection{Star Formation Rates}
\label{SFRs}
We characterized and compared the star formation rates of the
quasar host galaxies by comparing the
PAH emission with the far-IR luminosity of our sample (Fig.~\ref{PAHFIR}). We
measured the far-IR luminosity from only the modeled warm (Eq.~\ref{SGeq}) and cold dust (Eq.~\ref{FIReq}) contributions to the SEDs (hereafter L$'_{FIR}$), excluding dust heated by the central AGN
and emitting in the mid-IR. This definition of L$'_{FIR}$ specifically excludes dust heated by the AGN while still measuring dust heated in star forming regions, even if the two emit at similar or overlapping wavelengths. This is in contrast to using a defined wavelength cutoff, which would measure dust heated by either source. However, it is still important to note that the far-IR emitting dust could also be heated by the AGN if that dust were at large enough distances from the nucleus. It is important to consider this potential AGN contribution to the far-IR emission when calculating SFRs. 

The PAH emission is an independent indicator of the star formation activity in the host galaxy. UV and X-ray emission from the AGN destroy PAHs \citep{V92}, but soft UV emission from star formation excites PAHs into emission. In Fig.~\ref{PAHFIR} we find a correlation between PAH emission and far-IR emission similar to previous works (\eg~Schweitzer \etal~2006, Menendez-Delmestre \etal~2008). We also normalized the PAH emission to the far-IR emission, so we can compare the relative contribution of star formation to the far-IR emission. We measured the PAH luminosities by integrating our scaled templates, including the full profiles of each individual PAH feature in the template. This is different from many previous works (\eg~Schweitzer \etal~2006, Menendez-Delmestre \etal~2009), where only individual PAH features are measured, and definite wavelength cutoffs are applied. We find that the quasars as a whole have a fairly constant ratio across a range of far-IR luminosities. The median L$_{PAH}$/L$'_{FIR}$ ratio for the type-1 objects of our sample is 0.012, and the median for the type-2 objects is 0.021. We note that the difference between the type-1 objects and type-2 objects is not significant considering the scatter in the sample. The median of our overall sample is 0.017. In comparison, \citet{M07} measure PAH luminosity using a similar method to ours and find that the starburst galaxies NGC 7714 and NGC 2623 have L$_{PAH}$/L$_{FIR}$ ratios of 0.052 and 0.021, respectively. We cannot draw strong conclusions based on measurements of two objects, and note that many objects of our sample have ratios consistent with these objects. However, the average L$_{PAH}$/L$_{FIR}$ from Marshall \etal~is 0.037, higher than the median value of our sample. If taken at face value, these numbers indicate that some of the cool far-IR emitting dust is heated by AGN emission instead of stellar emission.

In Table~\ref{lum} we present the integrated mid- to far-IR emission ($8-1000$~\micron, hereafter L$_{IR}$). We find that  nearly all of our targets have log(L$_{IR}$/L$_{\odot}$) $\geq$ 11.75. SST 1721+6012 is the only object with log(L$_{IR}$/L$_{\odot}$) $\leq$ 11.75. We also list the star formation rates (SFR) of our sample in Table~\ref{lum}. We calculated SFRs in two different ways. First, we converted the integrated warm and cold dust components of our model into SFRs using the \citet{K98} calibration for starburst galaxies:  

\begin{equation}
SFR~[M_\odot~yr^{-1}] = 4.5 \times 10^{-44}L'_{FIR}~[erg~s^{-1}]
\end{equation}

Although our objects have IR luminosities similar to those of ULIRGs, their SFRs are lower than those of typical ULIRGs. This is because some of the IR luminosity of our objects is due to AGN-heated dust, and we explicitly exclude this component when calculating SFRs. The SFRs of our sample are instead typical of LIRGs, which can range from $20-200$ M$_\odot$ yr$^{-1}$ (\eg~Sanders \etal~2003). We find that the type-2 objects show higher SFRs on average (92 M$_\odot$ yr$^{-1}$) than the type-1 objects (67 M$_\odot$ yr$^{-1}$). However, the scatter is large, and the object with the lowest SFR is a type-2, SST 1721+6012. We also independently calculated the SFR from the PAH emission. This can be done by predicting a far-IR luminosity based on the PAH luminosity. We integrated our scaled PAH template to measure the PAH luminosity of each object in our sample. We adopt an average ratio of L$_{PAH}$/L$_{FIR}$ $\sim0.037$ based on the results of \citet{M07}. Subsequently, for all objects in our sample except two, the SFR predicted from the PAH luminosity is lower than the SFR predicted by integrating the warm and cold dust components of the far-IR (L$'_{FIR}$). Again we found that the average type-2 quasar has a higher SFR (87 M$_\odot$ yr$^{-1}$) than the average type-1 quasar (37 M$_\odot$ yr$^{-1}$).

\begin{deluxetable}{ccccc}
\tablecaption{Luminosities and Star Formation Rates}
\tablehead{ \colhead{Target} &
  \colhead{log(L$_{IR}$/L$_{\odot}$)$^{a}$} & \colhead{log(L$_{PAH}$/L$_{\odot}$)$^{b}$} &
  \colhead{SFR [M$_{\odot}$~yr$^{-1}$]$^{c}$} & \colhead{SFR [M$_{\odot}$~yr$^{-1}$]$^{d}$}}
\startdata
SDSS 171117.66+584123.8 & 12.25 & 9.3 & 109 & 10\\
SDSS 171126.94+585544.2 & 11.82 & 9.4 & 40 & 12\\
SDSS 171334.03+595028.3 & 11.95 & 9.3 & 27 & 10\\
SDSS 171736.91+593011.5 & 12.05 & 9.3 & 37 & 9\\
SDSS 171748.43+594820.6 & 12.06 & 10.3 & 66 & 98\\
SDSS 171818.14+584905.2 & 12.26 & 10.2 & 125 & 82\\
SST 171106.8+590436 & 11.79 & 10.0 & 73 & 48\\
SST 171147.4+585839 & 12.46 & 10.9 & 183 & 338\\
SST 171324.1+585549 & 12.13 & 10.0 & 85 & 52\\
SST 171831.7+595317 & 12.47 & 10.0 & 143 & 50\\
SST 172123.1+601214 & 11.48 & 9.0 & 14 & 4\\
SST 172458.3+591545$^{*}$ & 11.77 & 9.8 & 54 & 29\\
\enddata
\tablenotetext{a}{The luminostiy L$_{IR}$ corresponds to the integrated modeled SED from 8-1000~\micron.}
\tablenotetext{b}{The PAH luminosity is measured from the integrated profiles of all PAH features in the model.}
\tablenotetext{c}{Star formation rates derived from the sum of the warm and cold dust components of the model only. The SFR was calculated using the \citet{K98} relation.}
\tablenotetext{d}{Star formation rates derived from PAH luminosity using the results of \citet{M07} to convert from L$_{PAH}$ to L$_{FIR}$ and the \citet{K98} relation.}
\tablenotetext{*}{The fit of SST 1724+5915 underestimates the PAH luminosity and far-IR emission, thus also underestimating the SFRs derived from these quantities.}
\label{lum}
\end{deluxetable}

\section{Summary and Discussion}
We obtained mid-IR spectra and far-IR photometry for six type-1 and six type-2 mid-IR selected quasars. We modeled the SEDs of our sample from the optical to the far-IR using available SDSS, 2MASS and IRAC photometry, in addition to our IRS spectra and MIPS photometry. We compared the modeled SEDs of the quasars and calculated SFRs based on PAH and far-IR emission. Our main results are:

(1) The type-1 quasar mid-IR spectra show a featureless continuum with significant 3 \micron~emission. The full optical through far-IR SEDs are typically fit well with an optical quasar component, a 1000 K modified blackbody, a mid-IR power-law ($\sim3 - 30$\micron), and a cool dust far-IR  modified blackbody (45 K). Three of the objects also show evidence of warm small grain dust fit with a power-law ($\sim25 - 65$\micron).

(2) The type-2 quasar mid-IR spectra show a variety of PAH emission and silicate absorption strengths. The full optical through far-IR SEDs are fit well using a stellar population in the optical, a mid-IR power-law ($\sim3 - 30$\micron), varying amounts of a PAH emission template and silicate absorption in the mid-IR, warm small grain power-law emission ($\sim25 - 65$\micron) as well as a cool dust modified blackbody (45 K).

(3) Our averaged and normalized modeled spectra show that type-2 quasars exhibit more far-IR emission than type-1 quasars. The excess is roughly equivalent to the deficit in the optical with respect to the type-1 quasars, which is consistent with the orientation hypothesis. In this scenario the increased far-IR emission arises from obscuring dust that is reradiating absorbed light.

(4) We measured L$_{PAH}$ by integrating our scaled template for each object. We compared the typical L$_{PAH}$ to L$'_{FIR}$ ratio of our sample to that of starburst galaxies that have had L$_{PAH}$ measured in a similar manner as the method we used. Within our sample, we found that the type-2 objects have a larger ratio than the type-1 objects. We note that the difference between the two is not significant. Considering our overall sample, we found that the ratio is smaller than for the starburst galaxies \citep{M07}. This result suggests some far-IR emitting dust can be warmed by the central AGN rather than by young stars. However, a more direct comparison between AGN and starburst galaxies using larger samples would be needed to further quantify the amount of AGN-heated far-IR emitting dust.

(5) We calculated the SFRs of our sample using two methods. First we integrated the warm small grain and cool modified blackbody emission (excluding all other components) of the modeled SEDs and used the \citet{K98} relation to calculate the SFRs. We also converted the L$_{PAH}$ to L$_{FIR}$ \citep{M07} to measure the SFRs based on the PAH emission only. In each case the type-2 objects had higher SFR on average than the type-1 objects. However, the difference is not significant given the scatter.

(6) We integrated our modeled SEDs from $8 - 1000$~\micron~and found that seven of the 12 can be considered ULIRGs (log(L$_{IR}/$L$_{\odot}$) $\geq 12$), while four others have log(L$_{IR}$/L$_{\odot}$) $\geq 11.75$. The sole exception is SST 1721+6012, which has log(L$_{IR}/$L$_{\odot}$) $= 11.48$. While our sample appears to have IR luminosities similar to ULIRGs, the quasars exhibit SFRs more typical of LIRGs. We found the AGN can heat most of the dust in the mid-IR and even some in the far-IR. Thus measuring SFR in quasars from the full $8 - 1000$~\micron~SED overestimates the amount of dust warmed by young stars and the true SFR.

(7) The modeling results suggest that type-1 quasars can have mid-IR fluxes that have significant contributions from the optical power-law and 3 \micron~bump.  This 3 \micron~component is not required for the type-2 SEDs, and the hot dust most likely originates near the sublimation radius of the AGN. If this component is present in the type-2 quasars, it must be at least partially obscured by cooler dust, either a dusty torus or dust in the host galaxy. \citet{Lacy07} showed that the host galaxies of the type-2 quasars have a range of inclinations.  Therefore, the obscuration of the 3 \micron~bump, if it is intrinsically present, should be from material close to the nucleus and independent of galaxy orientation.

While our analysis does not distinguish definitively between orientation and evolution based unification schemes, it does directly address the dust content, ongoing star formation in the host galaxies and the AGN contribution to the far-IR emission. Our results are consistent with the orientation hypothesis, but we cannot rule out an evolutionary connection. The star formation based on PAH measurements is a larger contributor to the far-IR luminosity than the AGN for type-2 objects compared to type-1 objects. This difference is only suggestive due to the scatter of both L$_{PAH}$ and L$'_{FIR}$ in our sample and could arise from intrinsic differences in the bolometric luminosities of the objects. The L$_{PAH}$/L$'_{FIR}$ ratio in both type-1 and type-2 quasars is lower than that of starburst galaxies, suggesting that some emission in the far-IR is from dust heated by the AGN and not star formation. This is an area for continued research, as our analysis was based on few objects.

\acknowledgments
We thank the anonymous referee for useful comments and suggestions that helped improve both the paper's content and presentation. This work is based on observations made with the Spitzer Space Telescope, which is operated by the Jet Propulsion Laboratory, California Institute of Technology, under a contract with NASA. This research has made use of the NASA/IPAC Extraglactic Database (NED) which is operated by the Jet Propulsion Laboratory, California Institute of Technology, under contract with the National Aeronautics and Space Administration. Support for this work was provided by NASA through an award issued by JPL. Additional support was provided by the National Science Foundation, under grant number AST 0507450.

{\it Facilities:} \facility{Spitzer (IRS)}; \facility{Spitzer (MIPS)}

\end{document}